\documentclass[journal=jacsat,manuscript=article]{achemso}

\usepackage[version=3]{mhchem} 
\usepackage{subfig}
\usepackage{amssymb}
\usepackage[dvipsnames]{xcolor}

\author{J.~S.~Gomez-Diaz}
\email{juan-sebastian.gomez@utexas.edu}
\affiliation[Adaptive MicroNano Wave Systems, \'Ecole Polytechnique F\'ed\'erale de Lausanne (EPFL), 1015 Lausanne, Switzerland.]
{$\dag$ Adaptive MicroNano Wave Systems, \'Ecole Polytechnique F\'ed\'erale de Lausanne (EPFL), 1015 Lausanne, Switzerland.}
\alsoaffiliation[Department of Electrical \& Computer Engineering, University of Texas at Austin, 1 University Station C0803, 78712 Austin, Texas, USA.] {\\ $\ddag$ Department of Electrical \& Computer Engineering, University of Texas at Austin, 1 University Station C0803, 78712 Austin, Texas, USA.}
\author{C.~Moldovan}
\affiliation[Nanoelectronics Devices Laboratory, \'Ecole Polytechnique F\'ed\'erale de Lausanne (EPFL), 1015 Lausanne, Switzerland.]
{\\ $\P$ Nanoelectronics Devices Laboratory, \'Ecole Polytechnique F\'ed\'erale de Lausanne (EPFL), 1015 Lausanne, Switzerland.}
\author{S.~Capdevila}
\affiliation[Laboratory of Electromagnetics and Acoustics (LEMA), \'Ecole Polytechnique F\'ed\'erale de Lausanne, 1015 Lausanne, Switzerland.]
{\\ $\S$ Laboratory of Electromagnetics and Acoustics (LEMA), \'Ecole Polytechnique F\'ed\'erale de Lausanne, 1015 Lausanne, Switzerland.}
\author{J.~Romeu}
\affiliation[AntenaLAB, Universitat Politècnica de Catalunya, C/Jordi Girona 1-3, 08034 Barcelona, Spain]
{\\ $\|$ AntenaLAB, Universitat Politècnica de Catalunya, C/Jordi Girona 1-3, 08034 Barcelona, Spain}
\author{L.~S.~Bernard}
\affiliation[Laboratory of Physics and Complex Matter, \'Ecole Polytechnique F\'ed\'erale de Lausanne, 1015 Lausanne, Switzerland.]
{\\ $\bot$ Laboratory of Physics and Complex Matter, \'Ecole Polytechnique F\'ed\'erale de Lausanne, 1015 Lausanne, Switzerland.}
\author{A.~Magrez}
\affiliation[Laboratory of Physics and Complex Matter, \'Ecole Polytechnique F\'ed\'erale de Lausanne, 1015 Lausanne, Switzerland.]
{\\ $\bot$ Laboratory of Physics and Complex Matter, \'Ecole Polytechnique F\'ed\'erale de Lausanne, 1015 Lausanne, Switzerland.}
\author{A.~M.~Ionescu}
\affiliation[Nanoelectronics Devices Laboratory, \'Ecole Polytechnique F\'ed\'erale de Lausanne (EPFL), 1015 Lausanne, Switzerland.]
{\\ $\P$ Nanoelectronics Devices Laboratory, \'Ecole Polytechnique F\'ed\'erale de Lausanne (EPFL), 1015 Lausanne, Switzerland.}
\author{J.~Perruisseau-Carrier}
\affiliation[Adaptive MicroNano Wave Systems, \'Ecole Polytechnique F\'ed\'erale de Lausanne (EPFL), 1015 Lausanne, Switzerland.]
{$\dag$ Adaptive MicroNano Wave Systems, \'Ecole Polytechnique F\'ed\'erale de Lausanne (EPFL), 1015 Lausanne, Switzerland.}

\title[Self-biased Reconfigurable Graphene Stacks \\ for Terahertz Plasmonics]
{Self-biased Reconfigurable Graphene Stacks \\ for Terahertz Plasmonics}
\begin{document}

\begin{abstract}
The gate-controllable complex conductivity of graphene offers unprecedented opportunities for reconfigurable plasmonics at terahertz and mid-infrared frequencies. However, the requirement of a gating electrode close to graphene and the single `control knob' that this approach offers limits the practical implementation and performance of these devices. Here we report on graphene stacks composed of two or more graphene monolayers separated by electrically thin dielectrics and present a simple and rigorous theoretical framework for their characterization. In a first implementation, two graphene layers gate each other, thereby behaving as a controllable single equivalent layer but without any additional gating structure. Second, we show that adding an additional gate allows independent control of the complex conductivity of each layer within the stack and provides enhanced control on the stack equivalent complex conductivity. These results are very promising to the development of THz and mid-IR plasmonic devices with enhanced performance and reconfiguration capabilities.
\end{abstract}
\section{Introduction}
The strong graphene-light interaction has led to the rapid development of graphene plasmonics\cite{Koppens11, Grigorenko12}, which benefit from the unique electrical properties of graphene in the terahertz and mid-infrared frequency bands\cite{Geim2009}. The characterization of single-layer graphene structures has already been performed at microwaves\cite{Hao12,Sebas12_jap_measurements,Katsounaros13,Liang14}, terahertz\cite{Ren12,Buron12,Liang14} and optics\cite{Geim2009,Bao12}, and some promising applications such as modulators\cite{Ju11,Sensale-Rodriguez12,Sensale-Rodriguez12_nanoletters,Valmorra2013, DeglInnocenti2014,Lee12}, plasmonic waveguides\cite{Kim11_OptExpress,Christensen12} and Faraday rotators\cite{Crassee10} have been developed. However, the simple implementation and performance of these devices might be hindered by the presence of a gating electrode located close to graphene and the relatively weak control that this approach offers over the conductivity of graphene\cite{Ju11, Yan12}. These limitations can be overcome using graphene stacks, structures composed of two or more isolated graphene layers separated by electrically thin dielectrics, which lead to increased conductivity and may provide novel reconfiguration strategies.

Optical plasmons and quantum transport in such structures have already been studied theoretically\cite{Kechedzhi12, Stauber12, Zhu13}, whereas some experimental studies have focused on the Anderson localization of Dirac electrons in one of the graphene layers at DC due to the screening effect\cite{Ponomarenko11,Gorbachev12, Velasco12}. Furthermore, the Coulomb drag of massless fermions has been experimentally measured\cite{Kim2011}, while both intra- and inter- layer phenomena in structures surrounded by various dielectrics and their influence in the supported in-phase and out-of-phase plasmons have also been considered\cite{Badalyan2012,Fischetti2014}. Potential applications of graphene stacks include modulators\cite{Sensale-Rodriguez12}, enhanced metasurfaces\cite{Arya12}, antennas \cite{Tamagnone12_apl}, or plasmonic parallel-plate waveguides\cite{Christensen12, Diego13_mtt}, among many others. Experimentally, graphene stacks have recently been applied to the development of vertical FET transistors\cite{Schmidt08,Britnell12}. In addition, the response of \emph{unbiased} graphene stacks and devices at infrared frequencies has also been investigated\cite{Yan12}.

In this context, the work herein demonstrates the concept of \emph{reconfigurable} graphene stacks for THz plasmonics and presents a simple and rigorous theoretical framework for their characterization. Although the graphene monolayers within the stack are not close enough to couple through quantum effects\cite{Velasco12, Britnell12}, their extremely small separation in terms of wavelength allows the stack to behave as a single equivalent layer of increased conductivity. The enhanced tunable capabilities of the proposed structure are experimentally demonstrated in different scenarios, including the mutual gating between the graphene layers and the independent control of each sheet through two different biasing gates. The measurement of the total stack conductivity $\sigma_S$ \emph{for various combinations of gate voltages} permits not only the extraction of the different parameters that define each of the layers but also the determination of the effective gate capacitance of the surrounding dielectrics. The proposed formulation also allows the design of structures with the desired tunable conductivity behavior. Our results show that reconfigurable graphene stacks boost the available range of complex conductivity values provided by single-layer structures, thus facilitating the easy implementation of THz and mid-infrared plasmonic devices with enhanced reconfiguration capabilities.

\section{Results}

\subsubsection{Operation principle of reconfigurable graphene stacks.}
\begin{figure}[!h] \centering
\subfloat[]{\label{fig:_Artistic_double}
\includegraphics[width=0.5\columnwidth]{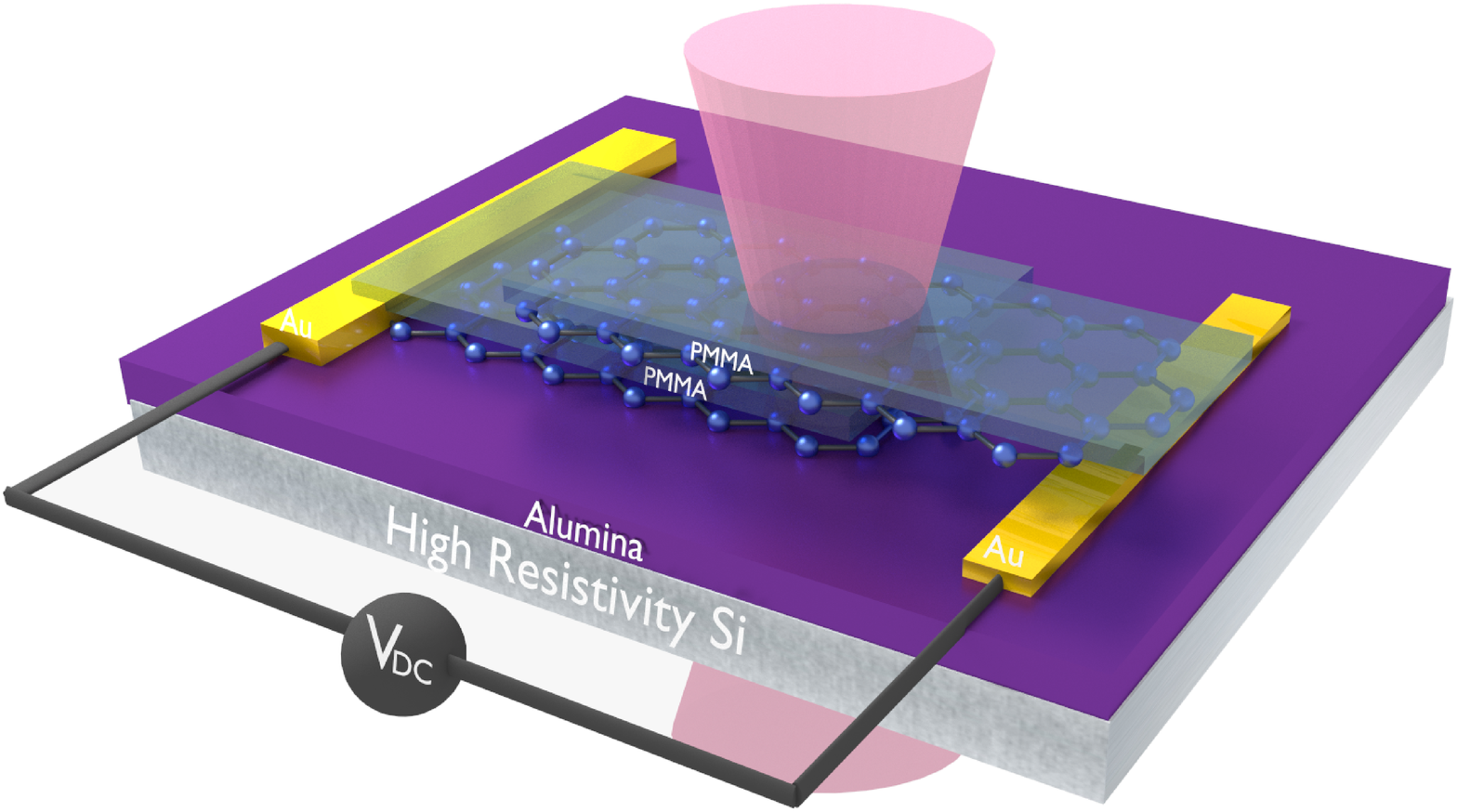}} \\
\subfloat[]{\label{fig:_Picture}
\includegraphics[width=0.3\columnwidth]{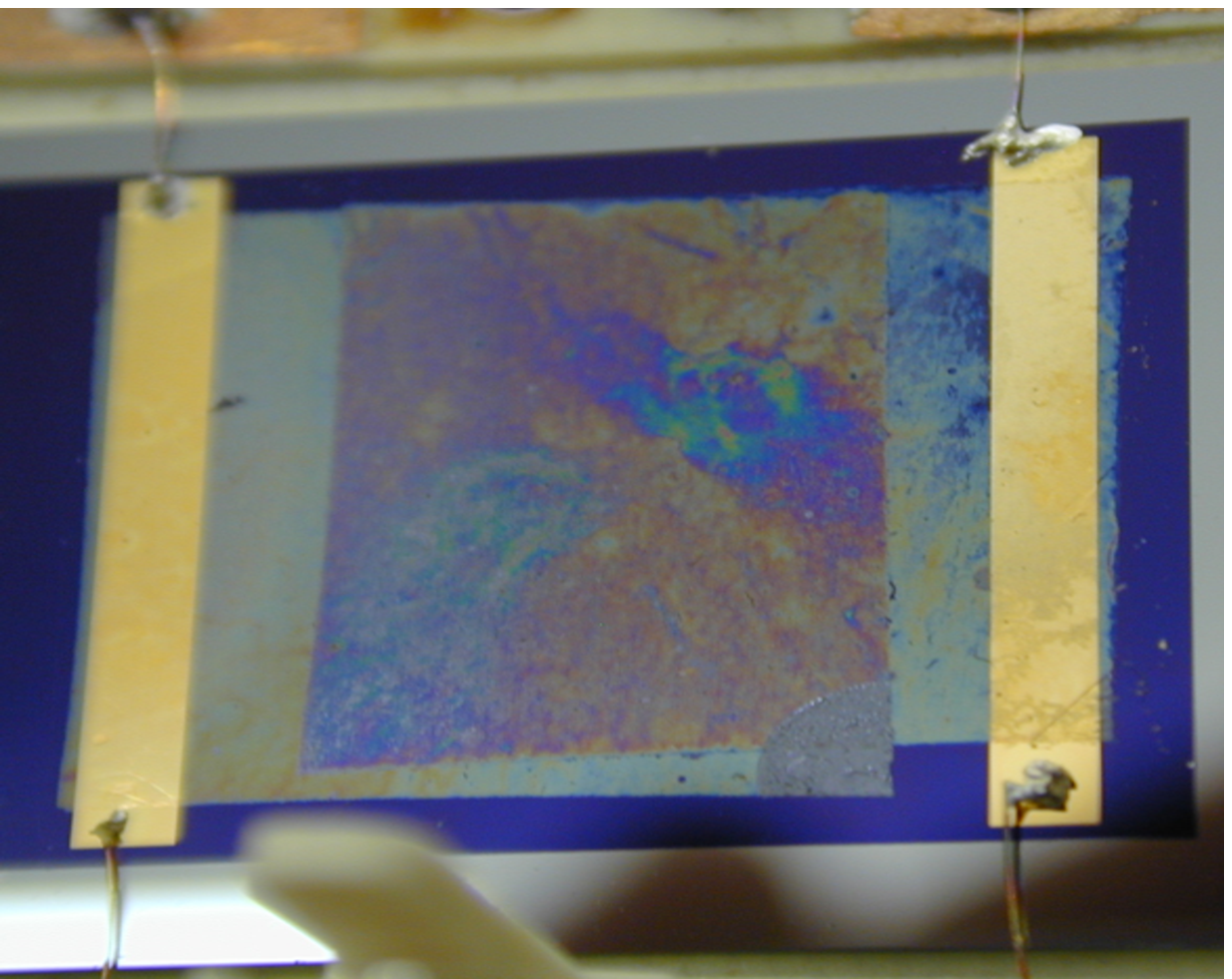}} \\
\subfloat[]{\label{fig:_Measurement_frequency}
\includegraphics[width=0.5\columnwidth]{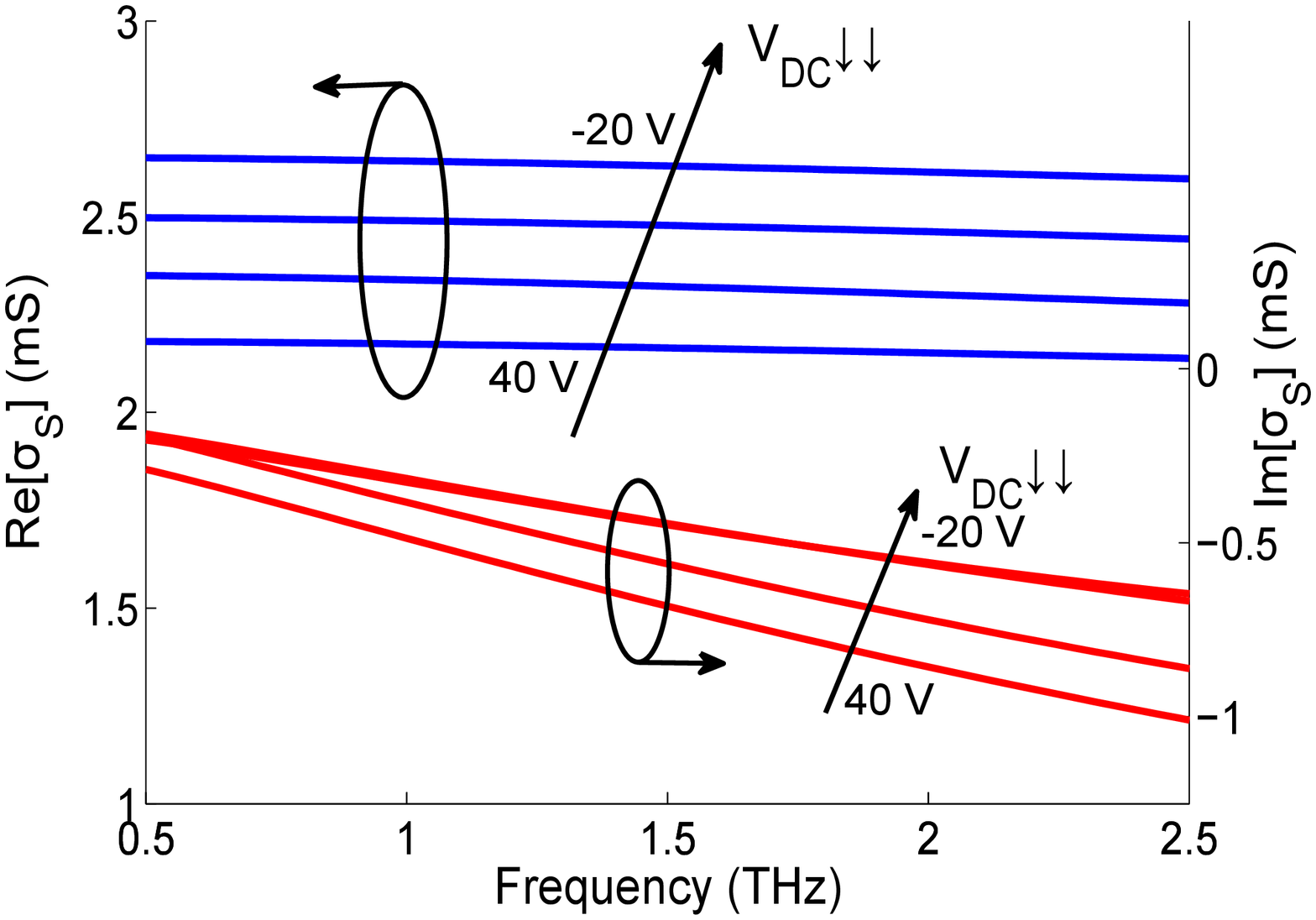}}
\caption{Graphene stack composed of two monolayer graphene sheets separated by an electrically thin layer of PMMA. (a) Artistic rendering of the fabricated sample. Incident and transmitted beams, employed for THz time-domain measurements, are illustrated for convenience. (b) Picture of the fabricated device. (c) Gate-controlled conductivity of the stack at terahertz. Measured real (blue) and imaginary (red) components of the conductivity are plotted versus frequency. Results are shown for various voltages $V_{DC}$ applied between the two graphene sheets of the stack.}
\label{fig:_Artistic}
\end{figure}

The structure under analysis is shown in Figure~\ref{fig:_Artistic_double}, where incident and transmitted beams required for THz time-domain measurements have been artistically rendered. The sample consists of two chemically vapor deposited (CVD) graphene monolayers separated by an electrically thin ($d \sim 80$~nm) polymethylmethacrylate (PMMA) layer. Metal contacts, added using optical lithography followed by the evaporation of $50$~nm of gold, have been included for biasing purposes. The sample is measured in the $0.5$-$2.5$~THz frequency range using time-domain spectroscopy. The complex conductivity of the graphene stack is then retrieved using a dedicated formulation\cite{Ren12, Crassee12, Naftaly07}. Details regarding the fabrication, measurement, and stack conductivity extraction process are provided in Methods. Because the dielectric separation layer between the graphene layers is extremely thin in terms of wavelengths ($d/\lambda_0\ll10^{-3}$) \cite{Yan12}, an incoming electromagnetic wave observes an stack conductivity $\sigma_{S}$
\begin{align}
\sigma_{S}=\sigma_{top}+\sigma_{bot}, \label{conductivit_stack}
\end{align}
where $\sigma_{top}$ and $\sigma_{bot}$ are the complex conductivity of the top and bottom graphene layers, respectively. Figure~\ref{fig:_Measurement_frequency} plots the frequency-dependent real and imaginary parts of the extracted conductivity $\sigma_{S}$ for several DC biasing voltages applied between the graphene layers. In the low terahertz band, the real component of the sample conductivity does not vary with frequency, whereas the imaginary part, which facilitates the propagation of surface plasmons in this frequency band\cite{Hanson08}, increases with frequency following a standard Drude model.

\begin{figure}[!t] \centering
\subfloat[]{\label{fig:_Meas_double_VDC}
\includegraphics[width=0.4\columnwidth]{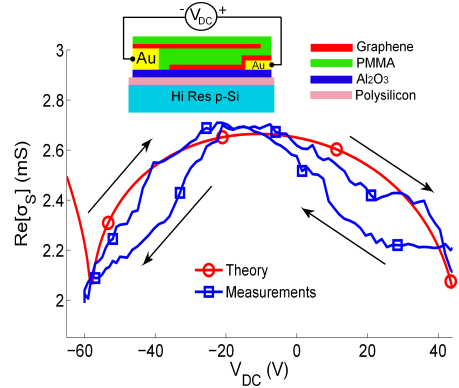}} \\
\subfloat[]{\label{fig:_Meas_double_2sources}
\includegraphics[width=0.4\columnwidth]{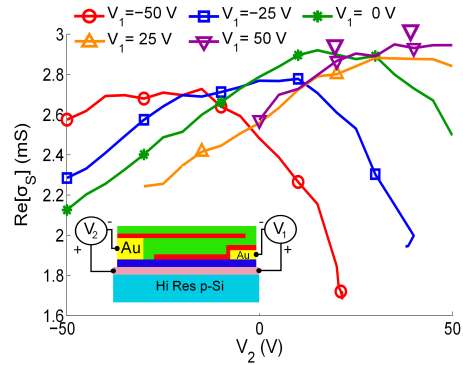}} \\
\subfloat[]{\label{fig:_Meas_double_2sources_difference}
\includegraphics[width=0.4\columnwidth]{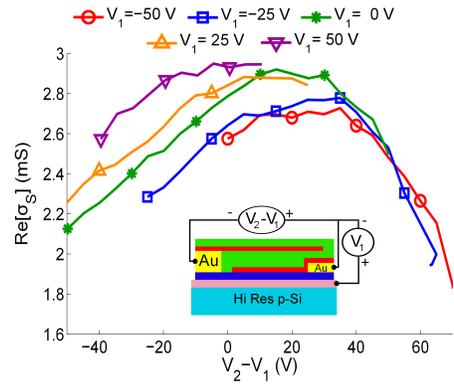}}
\caption{Reconfiguration capabilities of the fabricated graphene stack. (a) Measured conductivity plotted versus a voltage $V_{DC}$ applied between the two graphene sheets. Simulated results are included for comparison purposes. (b) Measured conductivity plotted versus the voltage $V_2$ applied to the top graphene sheet for different values of the bottom gate voltage $V_1$. (c) Measured conductivity plotted versus the voltage $V_2-V_1$ applied between the graphene layers for different values of the bottom gate voltage $V_1$. The different insets illustrate the sample cross-section and its connection to the voltage sources. For the sake of clarity, the hysteresis behavior of the stack conductivity has been removed in cases (b) and (c). The operation frequency is set to $f=1.5$~THz.} \label{fig:_Meas}
\end{figure}

\begin{figure}[!t] \centering
\includegraphics[width=0.75\columnwidth]{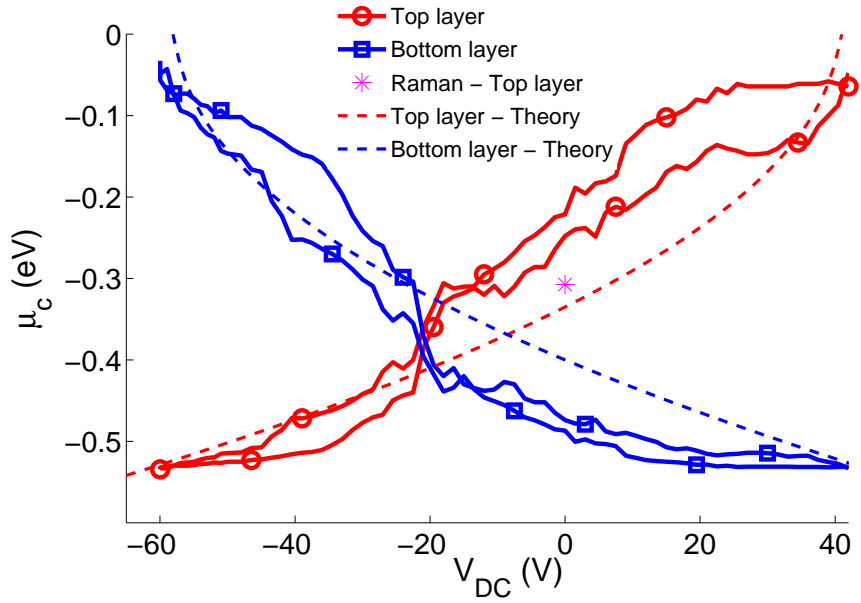}
\caption{Extracting the chemical potentials of the two layers composing the graphene stack. Results are computed versus the applied gate voltage $V_{DC}$ (see inset of Figure~\ref{fig:_Meas_double_VDC}), as detailed in Methods. Additional results obtained by measuring the dopants of the top layer using a Raman scattering technique\cite{Das08} (at $V_{DC}=0$~V) and by the proposed theory are included for comparison purposes.}
\label{fig:_Chemical_representation}
\end{figure}

Figure~\ref{fig:_Meas} shows the measured reconfiguration capabilities of the fabricated graphene stack at $f=1.5$~THz in different scenarios. In the first case, depicted in Figure~\ref{fig:_Meas_double_VDC}, a gate voltage $V_{DC}$ is applied between the two graphene layers. The results clearly confirm the tunability of $\sigma_S$ and the ability of the stack to self-bias. The extracted chemical potentials corresponding to each graphene sheet, computed using the procedure detailed in Methods combined with the measured stack conductivity and further validated by Raman scattering measurements\cite{Das08}, are depicted versus the applied gate voltage in Figure~\ref{fig:_Chemical_representation}. Both graphene layers are $p$-doped, and they present slightly different Fermi levels. This difference can be due to the defects induced in the graphene layers during growth or transfer\cite{Banhart11} and to the influence of the surrounding dielectrics\cite{Jang08,Konar10,Sharma14}. In addition to the different morphology of the surrounding dielectrics, contamination during processing\cite{Ishigami07} and molecules absorbed from ambient air\cite{Schedin07} play a crucial role. Applying a positive bias between the graphene sheets injects electrons/holes into the top/bottom layers, as illustrated in Supplementary Figure~$1$, which in turn increases/decreases their chemical potential (or vice versa in the case of a negative applied bias). Furthermore, the extracted conductivities and chemical potentials exhibit a hysteresis behavior, which arises due to the charges and impurities trapped in the surrounding dielectrics, as occurs in graphene transistors\cite{Wang10}.

Another interesting possibility for controlling the stack conductivity and boost its tuning range consists of applying voltages $V_1$ and $V_2$ to the bottom and top graphene layers, as illustrated in Figure~\ref{fig:_Meas_double_2sources}. For the sake of simplicity, we have implemented this biasing scheme by including an additional polysilicon gate below the lower graphene layer. Alternatively, this configuration might be implemented by stacking a higher number of graphene layers in the same structure. Finally, Figure~\ref{fig:_Meas_double_2sources_difference} presents a simple biasing procedure able to control the conductivity of each layer independently. Specifically, applying a fixed voltage $V_2-V_1$ between the graphene sheets fixes the chemical potential of the top layer, whereas the carrier density on the bottom layer is tuned by modifying the voltage $V_1$, as will be theoretically demonstrated below. Note that a voltage $V_1=0$~V does not exactly simplify this experiment to the one of Figure~\ref{fig:_Meas_double_VDC}, due to the weak electrostatic fields that may arise between the bottom graphene sheet and the polysilicon layers in practice (see Methods). The examples illustrated in Figure~\ref{fig:_Meas} demonstrate the large potential of graphene stacks for THz plasmonics, as it is possible to control the behavior of the different layers within a unique stack to achieve the complex conductivity required for a desired application.

\subsubsection{Static and dynamic characteristics.}
The graphene stack is theoretically analyzed in two different but interdependent steps. First, the carrier density on each graphene layer is determined as a function of the applied gate voltages using an electrostatic approach. Second, this information is employed to compute the frequency-dependent conductivity $\sigma_S$ of the stack. In a general case of two graphene sheets biased by different gate voltages $V_1$ and $V_2$ (see inset of Figure~\ref{fig:_Meas_double_2sources} and Supplementary Figure~$1$), these carrier densities can be approximated as \newline
\begin{align}
qn^{top}_s&=qn^{top}_{s_i}-C_{ox}^{top}\left(V_2-V_1\right), \label{density_top} \\
qn^{bot}_s&=qn^{bot}_{s_i}+C_{ox}^{top}\left(V_2-V_1\right)-C_{ox}^{bot}V_1, \label{density_bottom}
\end{align}
where $-q$ is the electron charge, $n^{p}_{s}$ is the total carrier density in the $p$ graphene layer (with p=$\{$bottom,top$\}$), $n^{p}_{s_i}$ corresponds to the pre-doping of the $p$ sheet, and $C_{ox}^{p}$ is the capacitance of the $p$ dielectric layer. Once the carrier densities are known, the Fermi level of each graphene layer and the conductivity $\sigma_S$, which determines the electromagnetic behavior of the whole stack, can be easily computed (see Methods for details). Moreover, Eqs.~\ref{density_top}-\ref{density_bottom} further confirm that it is possible to control the conductivity of each graphene layer independently. Specifically, the carrier density on both layers similarly depends on the difference between the applied voltages ($V_2-V_1$), while the bottom layer additionally depends on the voltage $V_1$. Consequently, modifying the voltage $V_1$ while keeping constant the difference $V_2-V_1$ allows the independent control of each layer's conductivity.

This simple framework allows a \emph{rigorous extraction of the characteristics of the stack from the measured data}, including the relaxation time ($\tau_{p}$) and the Fermi level ($\mu_{c}^{p}$) of the graphene layers and the capacitance ($C_{ox}^{p}$) of the surrounding dielectrics. This procedure, detailed in Methods, relies on applying different sets of gate voltages to the sample to measure various stack conditions, which in turns allows independent extraction of all of the aforementioned parameters. A system of nonlinear coupled equations is then imposed, relating the measured data to the theoretical characteristics of the stack. In the particular case of the sample shown in Figure~\ref{fig:_Meas_double_VDC}, the solution of the system of equations yields $\tau_{top}=0.033$~ps, $\tau_{bot}=0.03$~ps, $\mu_c^{top}=-0.4$~eV, and $\mu_c^{bot}=-0.355$~eV, whereas the gate capacitance of the PMMA separation layer is $C_{ox}^{top}=3.2\cdot10^{-4}$~F m$^-2$. These values are in good agreement with the measured characteristics of a single layer graphene transferred onto a similar dielectric ($\tau=0.029$~ps and $\mu_c=-0.425$~eV, as shown in the Supplementary Figures $2$-$4$) and with the gate capacitance obtained using the approximate parallel-plate formula $C_{ox}^{top}=\frac{\varepsilon_0\varepsilon_r}{d}\approx3.315\cdot10^{-4}$~F m$^-2$. The simulated results, plotted in Figures~\ref{fig:_Meas_double_VDC}-\ref{fig:_Chemical_representation} together with measured data, confirm the accuracy of both the extraction procedure and the proposed model to characterize reconfigurable graphene stacks. The measured hysteresis behavior of the sample conductivity, which is mainly related to the charges trapped in the dielectrics surrounding the graphene layers\cite{Wang10}, is not considered in the model. In addition, the extracted values permits estimating a modulation speed of $6.2$kHz for the stack (see Supplementary Note $1$), similar to the one found in single-layer graphene structures\cite{Sensale-Rodriguez12,Sensale-Rodriguez_13}.

\begin{figure} \centering
\subfloat[]{\label{fig:_2L_NP_same_real}
\includegraphics[width=0.5\columnwidth]{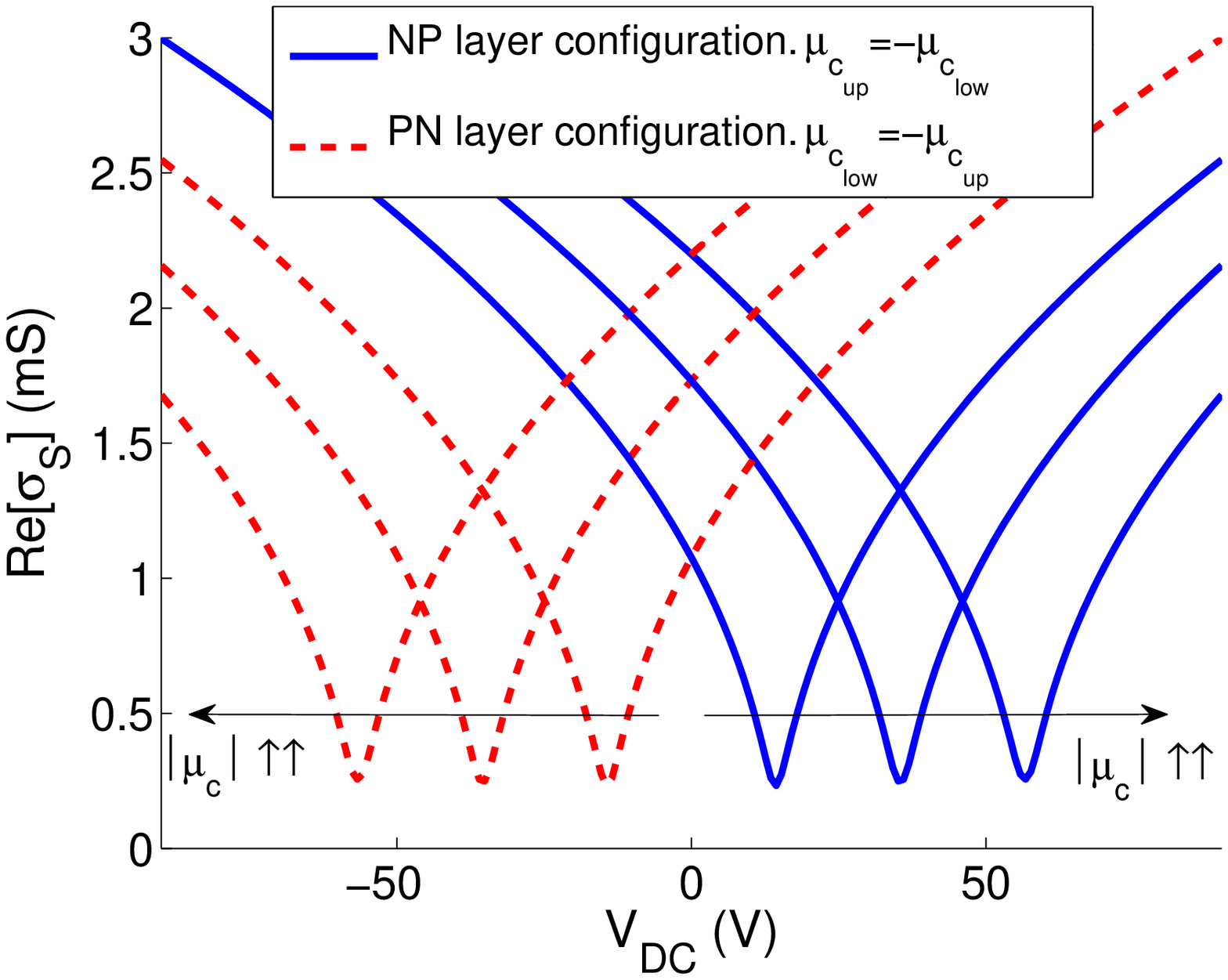}}
\subfloat[]{\label{fig:_2L_NP_same_imag}
\includegraphics[width=0.5\columnwidth]{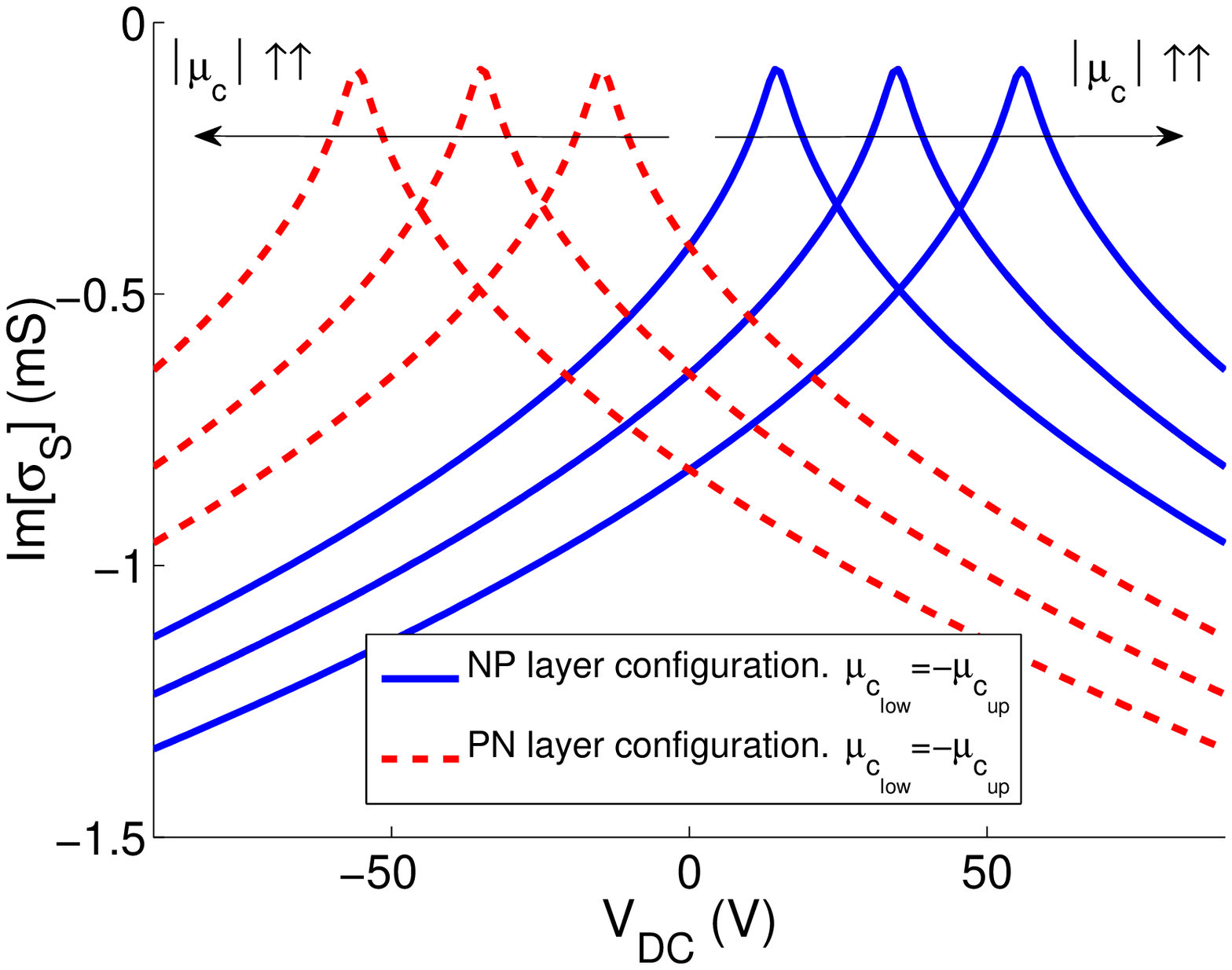}} \\
\subfloat[]{\label{fig:_2L_NN_same_real}
\includegraphics[width=0.5\columnwidth]{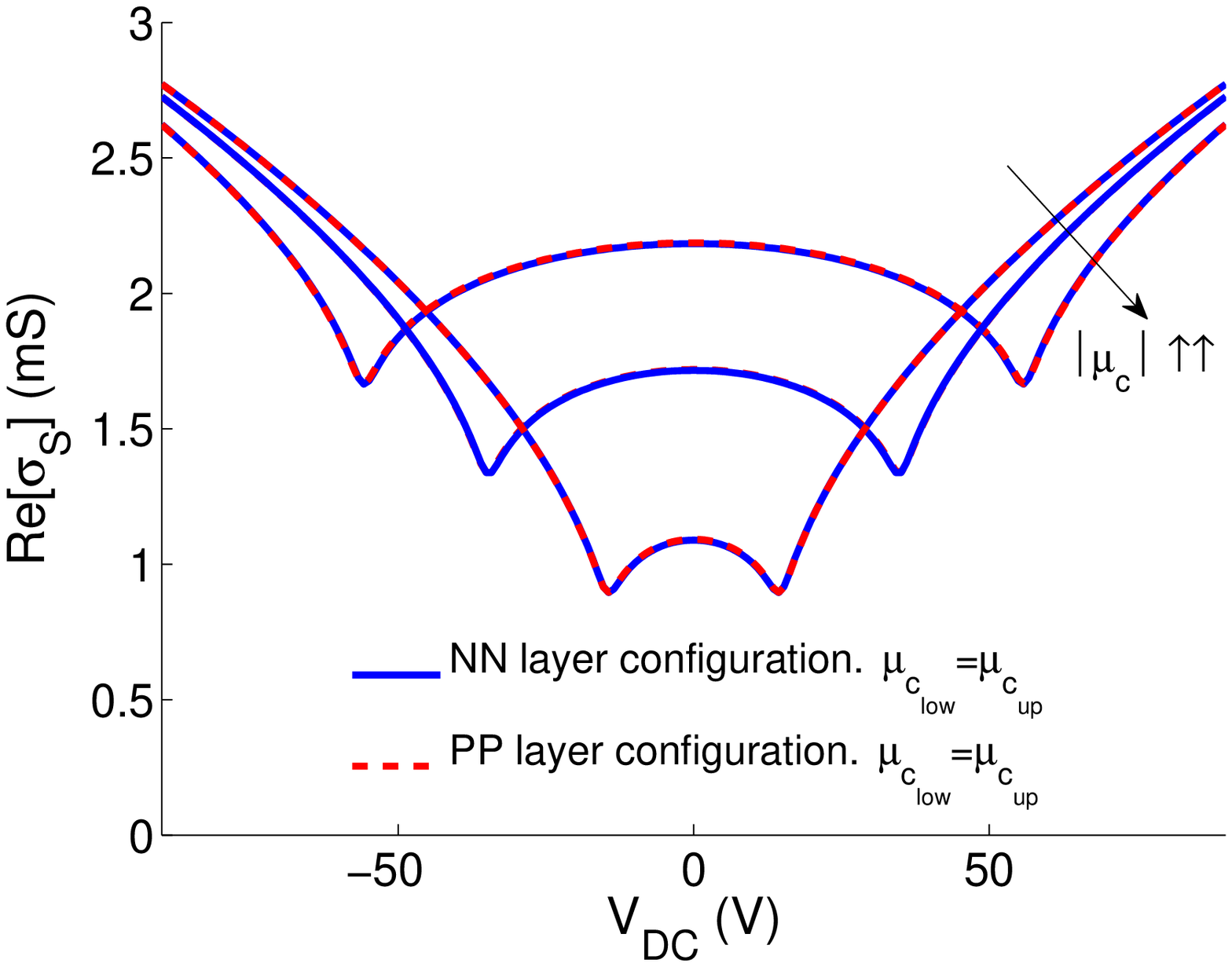}}
\subfloat[]{\label{fig:_2L_NN_same_imag}
\includegraphics[width=0.5\columnwidth]{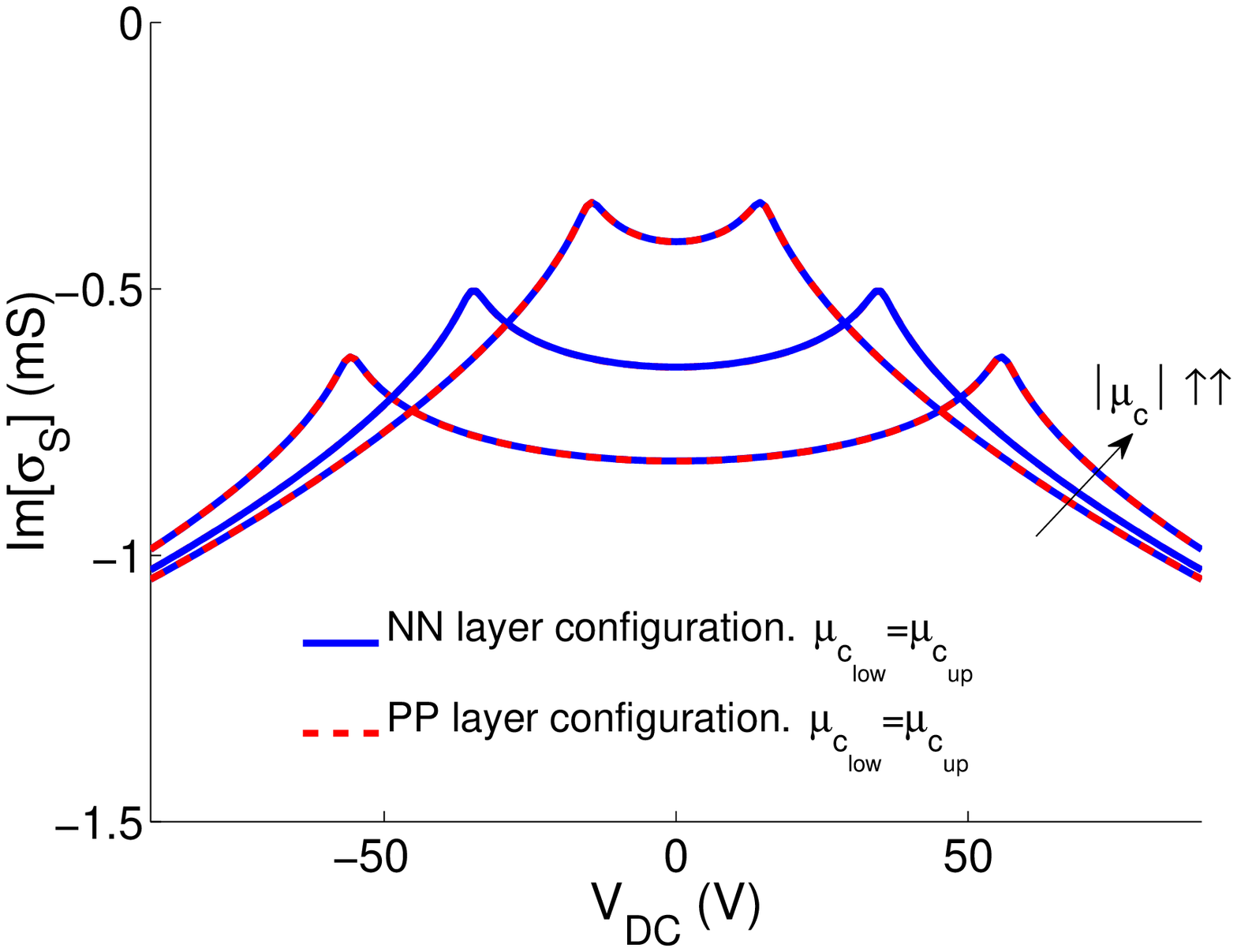}}
\caption{Theoretical reconfiguration capabilities of various graphene stacks. Results are computed versus the type of doping of the stacks composing layers. A biasing voltage $V_{DC}$ is applied between the graphene sheets, as illustrated in the inset of Figure~\ref{fig:_Meas_double_VDC}. The type of doping of the layers follows the nomenclature TB, where T=$\{$N,P$\}$ and B=$\{$N,P$\}$ are related to the top and bottom layers, respectively, and $\{$N,P$\}$ refers to $n$-doped or $p$-doped graphene. The upper row shows the real (a) and imaginary (b) conductivity components of a stack composed of layers with opposite Fermi level. (c)-(d) shows similar results for the case of a stack composed of layers with equal Fermi level. Other parameters are $f=1.5$~THz, $\tau_1=\tau_2=0.03$~ps and $T=300$~K.}
\label{fig:_2L}
\end{figure}

This framework can be further employed to forecast the reconfiguration capabilities of a wide variety of graphene stacks, allowing the design of structures with desired plasmonic properties and tunable behavior. Figure~\ref{fig:_2L} illustrates at $f=1.5$~THz the real and imaginary conductivity components of a stack composed by two graphene sheets with various Fermi levels versus the gate voltage applied between the layers (see inset of Figure~\ref{fig:_Meas_double_VDC}). Let us first consider, for simplicity, a stack where the graphene layers have a different type of doping, i.e., one sheet is $p$-doped and the other is $n$-doped. In this particular case, illustrated in Figsures~\ref{fig:_2L_NP_same_real}-\ref{fig:_2L_NP_same_imag}, the carriers injected by the voltage source alter the carrier density on each layer in a similar way, i.e., simultaneously increasing/decreasing their $\mid\mu_c^{p}\mid$ while keeping their opposite doping nature $\left[\text{sign}(\mu_c^{top})\neq \text{sign}(\mu_c^{bot})\right]$. As a result, the stack conductivity is approximately twice the conductivity of an individual layer. The behavior of the stack conductivity differs with respect to the previous case when the layers have the same type of doping, i.e., if they are both $p$-doped or $n$-doped. In this case, shown in Figures~\ref{fig:_2L_NN_same_real}-\ref{fig:_2L_NN_same_imag}, the carriers injected by the source modify the carrier density on each graphene sheet in an opposite direction, i.e., increasing $\mid\mu_c\mid$ of one layer while reducing it on the other layer. Consequently, the stack conductivity presents a symmetrical behavior for positive and negative gate voltages, exhibiting points of minimum conductivity in both cases. The results shown in Figure~\ref{fig:_2L} confirm that in graphene stacks i) the imaginary component of $\sigma_S$ can be double than the one of a single-layer structure, while avoiding the presence of metallic bias, and ii) the tuning range is significantly boosted for similar applied voltages values. In addition, the conductivity of graphene stacks can be controlled further by considering two different gate sources, as shown in the inset of Figure~\ref{fig:_Meas_double_2sources}. Similar to the previous case, the tunable behavior of the stack conductivity will strongly depend on the initial level and the type of doping of each graphene layer, leading to a wide variety of scenarios and reconfiguration possibilities (see Supplementary Figures $5$-$9$).

\subsubsection{Surface plasmons supported by graphene stacks.}
\begin{figure}[!h] \centering
\subfloat[]{\label{fig:_PPW_Even_mode}
\includegraphics[width=0.5\columnwidth]{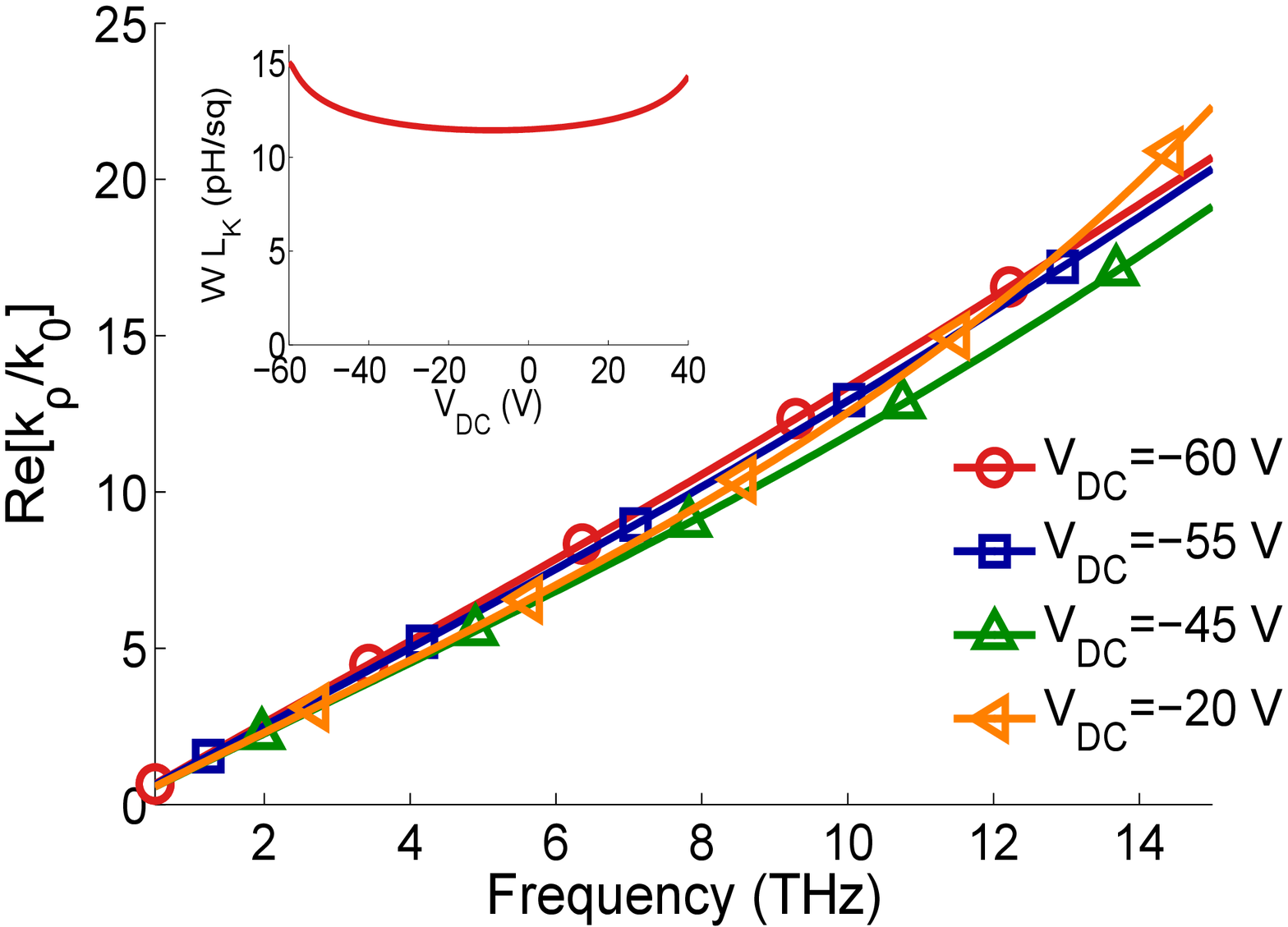}}
\subfloat[]{\label{fig:_PPW_Odd_mode}
\includegraphics[width=0.5\columnwidth]{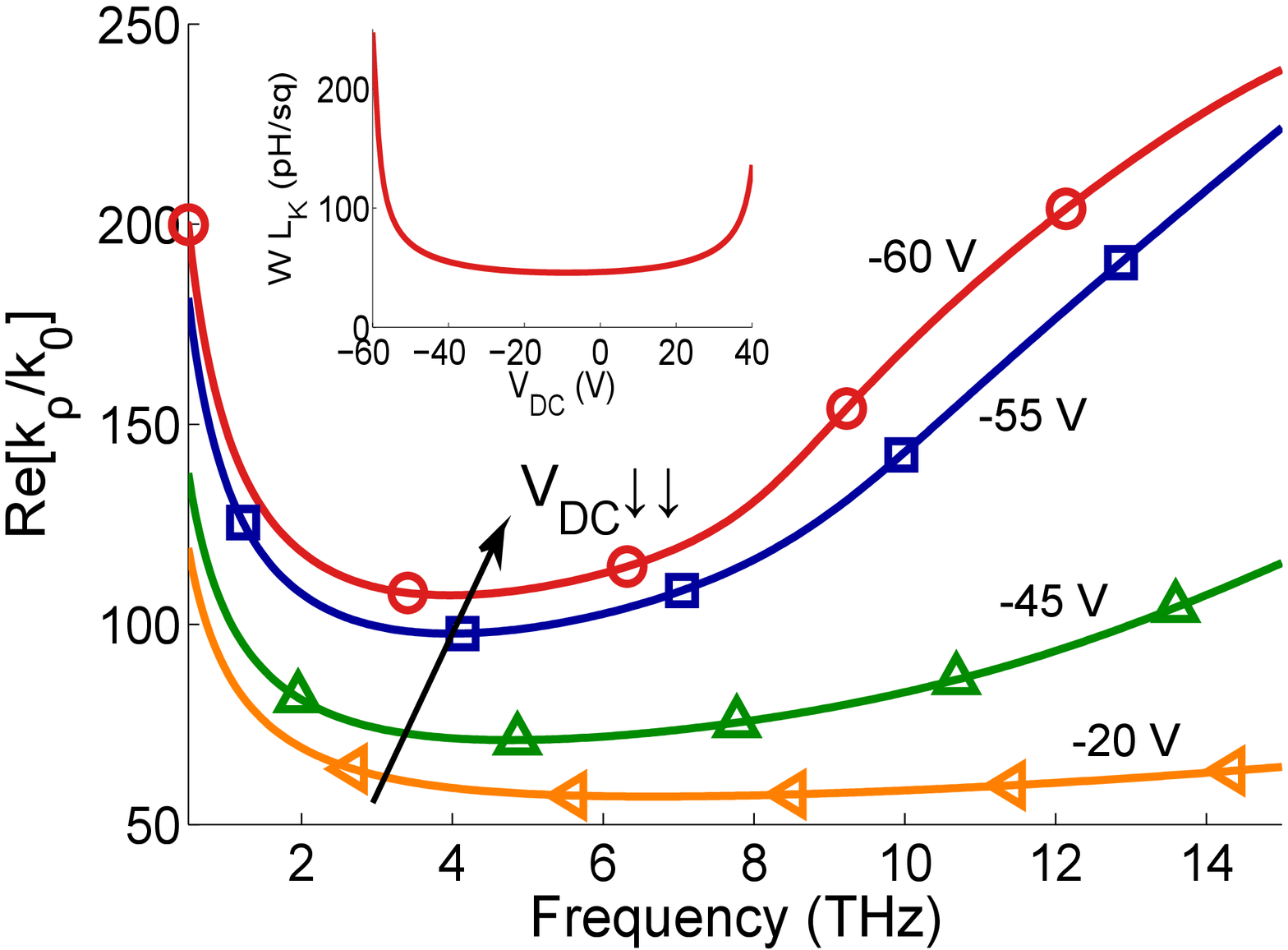}}
\caption{Characteristics of the surface plasmons supported by the fabricated graphene stack. Results are shown for several values of the voltage V$_{DC}$ applied between the two graphene sheets of the stack (see inset of Figure~\ref{fig:_Meas_double_VDC}). (a) TM mode. (b) Quasi TEM mode. The insets show the kinetic inductance associated to each mode versus the applied bias assuming a width $W=10\mu m$. Simulations (see Methods) are performed using the extracted characteristics of the stack as input parameters.}
\label{fig:_PPW_modes}
\end{figure}

The measured characteristics of the fabricated stack allows to simulate the frequency-dependent properties of the surface plasmons supported by the device. Specifically, the structure supports two different modes\cite{Hanson08_PPW,Diego13_mtt} (see Methods): an even TM and an odd quasi-TEM. The former can easily be seen as a usual TM plasmon propagating along a single-layer graphene sheet with a conductivity equal to the stack conductivity $\sigma_S$. Figure~\ref{fig:_PPW_Even_mode} illustrates the characteristics of this mode, which presents lower field confinement and reduced tunability compared to plasmons in single-layer graphene. This behavior arises due to the increased imaginary component of the stack conductivity, which in turn reduces the kinetic inductance associated to this mode. The later is a perturbation of the TEM mode found in standard parallel-plate waveguides with two perfect electric conductors. Figure~\ref{fig:_PPW_Odd_mode} confirms that this mode presents remarkable characteristics in terms of field confinement and tunability, clearly outperforming single-layer graphene structures. Note that the high losses associated to CVD graphene\cite{Song12}, which prevent the propagation of the supported plasmons along many wavelengths, can be significantly mitigated employing high-quality graphene in the stack\cite{Yoon14}. Supplementary Note $2$ includes a comparison of the characteristics of plasmons supported by the stack and a single-layer graphene structure, and further discusses the influence of losses in both cases.

\section{Discussion}
This theoretical and experimental study of graphene stacks has demonstrated that the available range of complex conductivities in graphene stacks can be significantly boosted by two different approaches i) mutually biasing the graphene sheets without requiring the presence of any metallic bias, and ii) including a third gate source to control the conductivity of each layer independently. The development of graphene stacks for terahertz plasmonic also faces some important challenges from the technological point of view, as it would be desirable to independently control the doping nature of each layer while decreasing the separation distance between the graphene sheets in order to further enhance the reconfiguration possibilities of the stack. Furthermore, inter-layer Coulomb effects\cite{Badalyan2012,Fischetti2014} should be rigorously taken into account in stacks with very small ($\sim$nm) separation distances between their layers. In addition, it would be also interesting to extend the concept of reconfigurable stacks to an arbitrarily large number of layers. The exotic characteristics of graphene stacks paves the way towards the development of a low-dimensional plasmonic platform with enhanced performance and reconfiguration capabilities. For instance, graphene stacks are the building block of the recently proposed tunable bulk hyperbolic metamaterials\cite{Othman_13}, whereas it has also demonstrated that they are able to boost of the spontaneous emission of emitters\cite{Zhang13_OptExpress} much further than usual monolayer graphene structures\cite{Koppens11}. In a different context, the large range of imaginary conductivity values provided by the stacks can easily be exploited in planar hyperlenses. Currently, graphene-based hyperlenses\cite{Forati14} are based on achieving large contrast of conductivities within the surface by using non-uniform metallic gates located very close to graphene. However, these gates are difficult to fabricate and impair the performance of the lens. This device could easily be implemented by a patterned graphene stack, simultaneously solving the problems related to the limited values of the imaginary conductivity and the presence of the nonuniform metallic gate. Finally, note that the aforementioned features of graphene stacks can also be applied to develop improved devices such as modulators, isolators, sensors, or antennas in the THz and infrared frequency bands.

\section{Methods}
\subsubsection{Fabrication of single-layer and stack graphene structures.}
The samples were fabricated using CVD graphene grown on Cu foil and transferred onto the substrate using the standard wet transfer method \cite{Reina2008}.  Supplementary Figure~$10$ shows the flow of the fabrication process for the double-layer graphene stack. We deposit $72$nm of Al$_2$O$_3$ by atomic layer deposition (ALD) on a an ultrahigh-resistivity ($>10\text{k}\Omega$) p-type Si wafer. The ALD is performed at $200^{\circ}$~C using trimethylaluminum (TMA) and distilled water as the reaction precursors. Prior to the dielectric deposition, the native oxide is removed from the Si wafer with a buffered oxide etch. The metal electrodes are patterned by optical lithography followed by a deposition of $5$~nm of chromium, $50$~nm of gold and a lift-off process. A graphene sheet is then transferred onto the top of one of the metal contacts.  In the double-layer graphene stack, the polymethylmethacrylate (PMMA) layer used as a support polymer during the transfer process is kept on top of the graphene to act as a dielectric between the two graphene sheets. The second graphene monolayer is subsequently transferred onto the top of the other predefined metal contact, thus obtaining the final structure shown in Supplementary Figure~$10$d.

The Raman spectra of the graphene employed in our devices is shown in Supplementary Figure~$11$. The G and $2$D band points are located at $1589$ and $2682$ cm$^{-1}$ with a full width at half maximum of $18$ and $32$ cm$^{-1}$, respectively. The intensity ratio of the $2$D to the G band and of the D to the G band are $5.5$ and $0.09$. All of these numbers are typical hallmarks of monolayer graphene\cite{Ferrari06}. The use of spincoated PMMA as a separation layer between the graphene sheets allows the avoidance of problems associated with standard dielectric deposition techniques such as evaporation, sputtering\cite{Lemme07} and ALD of oxides, which can induce defects in graphene. This approach is convenient for fabricating graphene stacks, allowing viable biasing schemes without the need of post-processing the graphene. Note that the DC isolation between the two graphene layers of the fabricated stack is not perfect, and some leakage current has been measured. However, it does not hinder the performance of the stack since i) the device does not operate at DC but in the THz band, and ii) graphene field's effect control is preserved as the DC biasing voltage source is able to provide the required bias voltage, hence the required electrical field,  even when some leakage current occurs.

In addition, note that monolayer graphene devices have been annealed in a N$_2$ atmosphere at $200^{\circ}$C during $4$ hours \cite{Xueshen13}. The annealing aims removing possible graphene contamination by polymer residues and other impurities \cite{Lin12Annealing}. However, this process has not been applied to the graphene stack samples because it would remove the PMMA layer which isolates the two graphene sheets.

\subsubsection{THz time-domain measurements.}
The measurements at terahertz frequencies were performed using a commercial Time Domain Spectrometer (TDTS) (Menlo TERA-K8), which consists of a pulsed femtosecond laser at $780$~nm with a pulse repetition rate of $100$~MHz and a pulse width around $110$~fs, offering with the current experimental setup a signal-to-noise ratio (SNR) of $40$~dB up to $2.5$~THz. Two photoconductive antennas based on LT GaAs (Tera$8$-$1$) are used to generate and detect the THz radiation. A set of lenses focuses the THz beam onto the sample under measurement. The total sample area illuminated by the beam is around $2$~mm$^2$, thus averaging graphene's features (see Supplementary Note $3$). Supplementary Figure~$12$ presents a schematic view of the experimental setup, with the disposition of the samples and voltage sources.


The gating was applied using a $4$-channel DC voltage source, Agilent N$6700$B. Only $2$ channels were used for the measurements, and each channel was connected to a different gold contact corresponding to a graphene layer, whereas they both shared a common ground gold contact. For safety reasons and to prevent damaging the graphene stack, the maximum voltage (taking into account both sources) was limited to $\pm 75$~V. The sample was placed on an X-Y linear stage perpendicular to the THz beam, and everything was placed inside a sealed case purged with $N_2$ to keep a constant atmosphere during the duration of the measurements.

\subsubsection{Stack conductivity extraction.}
The stack conductivity $\sigma_S$ is extracted from the THz time-domain measurements using standard thin-film characterization techniques\cite{Nuss1998,Naftaly07,Ren12,Crassee12}. This approach is valid here thanks to the extreme fineness of the stack in terms of wavelength ($d/\lambda_0\ll10^{-3}$). An example of the different set of measured pulses employed for the extraction procedure is shown in Supplementary Figure~$13$. To keep the higher possible SNR, we have considered only the first transmitted pulse through the sample. Additional transmitted pulses that arise due to the internal reflections of the THz beam within the layers of the sample are clearly identified thanks to their temporal delay and subsequently removed.

The graphene stack is not free-standing but on top of a thick dielectric structure. Consequently, the influence of the dielectrics must be rigorously removed to extract the actual stack conductivity. This procedure has been performed as follows: i) A pulse is transmitted without the presence of any sample to measure and store the response (including atmosphere and possible impurities) of the sealed cage. ii) A pulse is transmitted through an area of the sample free of graphene, which remains bare. The combination of this measured pulse with the pulse obtained in the previous step allows the extraction of the permittivity, loss tangent and thickness of the dielectrics using standard techniques\cite{Nuss1998,Naftaly07}. iii) A pulse is transmitted through the graphene stack sample. Combining this measured pulse with the previous information, it is indeed possible to extract the conductivity of the graphene stack rigorously removing the influence of the dielectrics and surrounding atmosphere\cite{Naftaly07,Ren12,Crassee12}.

\subsubsection{Graphene stack theory.}
The frequency-dependent conductivity $\sigma$ of a single graphene layer is modeled using the Kubo formalism\cite{Gusynin09} as
\begin{align}
\sigma(\omega,\mu_c,\Gamma,T)=&\frac{i q^2(\omega-i2\Gamma)}{\pi \hbar^2}\left[\frac{1}{(\omega-i2\Gamma)^2} \int_0^{\infty} \epsilon \left(\frac{\partial f_d(\epsilon)}{\partial \epsilon}-\frac{\partial f_d(-\epsilon)}{\partial \epsilon}\right)\partial\epsilon- \right. \nonumber \\ & \left. \int_0^{\infty} \frac{f_d(-\epsilon)-f_d(\epsilon)}{(\omega-i2\Gamma)^2-4(\epsilon/\hbar)^2}\partial\epsilon\right], \label{eq: Conductivity}
\end{align}
where $\omega$ is the radian frequency, $\epsilon$ is energy, $\Gamma=1/(2\tau)$ is a phenomenological electron scattering rate assumed independent of energy, $\tau$ is the electron relaxation time, $T$ is temperature, $-q$ is the charge of an electron, $\hbar$ is the reduced Planck's constant, and $f_d$ is the Fermi-Dirac distribution defined as
\begin{equation}
f_d(\epsilon)=\left(e^{(\epsilon-\mu_c)/k_BT}+1\right)^{-1},
\label{eq_Fermi_distribution}
\end{equation}
being $\mu_c$ the chemical potential and $k_B$ Boltzmann's constant. This model results from the long wavelength limit of the bosonic momentum ($k_{\mid\mid}\rightarrow 0$) and takes into account both intraband and interband contributions of the graphene conductivity as well as a finite temperature.

In addition, the carrier density $n_s$ and chemical potential of the graphene layer are related through
\begin{equation}
n_s=n_{s_e}-n_{s_h}=\frac{2\text{sign}(\mu_c)}{\pi\hbar^2v_f^2}\int_0^\infty\epsilon[f_d(\epsilon-\mu_c)-f_d(\epsilon+\mu_c)]\partial\epsilon,
\label{eq:Carrier_densitiy}
\end{equation}
where $n_{s_e}$ and $n_{s_h}$ are the electron and hole densities, respectively, $\epsilon$ is energy and $v_f$ is the Fermi velocity ($\sim 10^8$ cm s$^{-1}$ in graphene).

Let us consider the case of two graphene layers closely located within a stack, as depicted in Supplementary Figure~$1$. As previously stated, the structure is analyzed first using an electrostatic approach, which determines the carrier density on the graphene layer, and then obtaining the electromagnetic behavior of the stack at THz. Following the superposition principle (see Supplementary Figure~$1$b), the carrier density on each layer are computed using Eq. (2)-(3). Note that this electrostatic approach approximates both graphene and polysilicon for infinite perfect conductors in order to compute the carrier density on each layer. Consequently, it cannot predict the presence of weak electrostatic fields that may arise due to i) the different DC conductivities that graphene and polysilicon present in practice, and ii) fringing effects at graphene borders. Combining these expressions with Eq.~(\ref{eq:Carrier_densitiy}) permits the chemical potential on each graphene layer to be determined. Once these potentials are known, the frequency-dependent complex conductivity of the individual graphene sheets is retrieved using Eq.~(\ref{eq: Conductivity}), thus allowing the total graphene stack conductivity to be computed using Eq.~(\ref{conductivit_stack}). Note that in this approach we have neglected i) the influence of the separation layer located between the graphene sheets, which is electrically very small in the THz frequency range, and ii) the possible influence of the quantum capacitance\cite{Chen08}, which may be significant in the case of high permittivity or extremely thin ($\sim$nm) dielectrics but is completely negligible here.

Note that the proposed approach approximates graphene's relaxation time as a constant quantity in each layer, and embeds all variations of graphene conductivity versus the applied bias in the chemical potential\cite{Liang14}, \cite{Liu14},\cite{Ren12}. However, rigorous approaches indicate that the relaxation time not only depends on the defects in graphene ($\tau_{gr}$), but also on the thermally excited surface polar phonons that may arise at the interface between graphene and the substrate ($\tau_{sb}$), and on the frequency-dependent electron-phonon coupling ($\tau_{e-ph}$) \cite{Jablan09}. These values are related through the Matthiessen's rule \cite{Zhu09} by $\tau^{-1}=\tau_{gr}^{-1}+\tau_{sb}^{-1}+\tau_{e-ph}^{-1}$. In addition, graphene relaxation time and chemical potential are not totally independent\cite{Jablan09}. In our particular experiments, the extracted relaxation times are very similar. Since the operation frequency is in the low THz range, well below the graphene optical phonon frequency \cite{Jablan09}, we expect that electron-phonon phenomenon does not impact the $\tau$ decay mechanism. The similarities among the extracted relaxation times, which correspond to graphene layers surrounded by different substrates, suggests that the graphene/dielectric interface provides a high $\tau_{sb}$ thus being graphene impurities and nonidealities ($\tau_{gr}$) the main mechanism limiting the relaxation time, i.e.  $\tau\approx\tau_{gr}$. Other possible effects such as carrier scattering by ionized impurities \cite{Adam07} and the electron-hole puddle effect \cite{Chen08b} might also modify the measured relaxation time.
\subsubsection{Extracting the characteristics of each graphene layer.}
Let us consider a stack composed of two graphene layers, which are biased by two different gate sources as depicted in Supplementary Figure~$1$. The availability of the total stack conductivity $\sigma_S$ \emph{for various combinations of gate voltages} not only permits the extraction of the different parameters which define each of the layers but also determines the effective gate capacitance of the surrounding dielectrics. Specifically, given a measured stack conductivity $\sigma_{S}^{i}$ obtained by applying a set $i$ of gating voltages ($V_1^{i}$, $V_2^{i}$), Eq.~\ref{conductivit_stack} holds. This equation relies on the top and bottom complex conductivities of the layers ($\sigma_{top}^{i}, \sigma_{bot}^{i}$), which are computed using the Kubo formula of Eq.~\ref{eq: Conductivity} and depend on their relaxation time ($\tau_{top}$, $\tau_{bot}$), Fermi levels ($\mu_{c}^{top}$, $\mu_{c}^{bot}$), gate capacitance of the surrounding media ($C_{ox}^{top}$, $C_{ox}^{bot}$), and the applied gate voltages. Considering a set of $N$ measured stack conductivity values, obtained by applying different gate voltages, permits the extension of Eq.~\ref{conductivit_stack} into a set of nonlinear coupled equations. The numerical solution of this system of equations determines both the characteristics of each graphene layer within the stack and the gate capacitance of the surrounding media. The solution of these equations may differ slightly as a function of the measured conductivity data employed as an input. These small variations are related to diverse factors, including the hysteresis of the stack conductivity\cite{Wang10} or the possible change in the environmental conditions\cite{Moser08} (especially humidity) during the measurements. In order to take them into account, the various parameters extracted from all possible combination of gate voltages are finally averaged. Employing a curve fitting approach is not straightforward here as this would involve fitting $6$ independent variables, which could lead to non-physical parameter values and complicated post-processing steps.

\subsubsection{Surface plasmons supported by a graphene stack.}
The dispersion relation of the plasmonic modes supported by a graphene stack can be computed as\cite{Hanson08_PPW}
\begin{equation}
(c_{top}+c_{bot})\cos(k_{z_2}d)+i(c_{top}c_{bot}+1)\sin(k_{z_2}d)=0
\label{eq_PPW_modes}
\end{equation}
where
\begin{align}
c_{top}&=\left(\frac{\varepsilon_2k_{z_1}}{\varepsilon_1k_{z2}}\right)\left(1+\frac{\sigma_{top}k_{z_1}}{\omega\varepsilon_1\varepsilon_0}\right)^{-1}, \\
c_{bot}&=\left(\frac{\varepsilon_2k_{z_3}}{\varepsilon_3k_{z2}}\right)\left(1+\frac{\sigma_{bot}k_{z_3}}{\omega\varepsilon_3\varepsilon_0}\right)^{-1},
\label{eq_PPW_modes_inside}
\end{align}
$k_{z_i}^2=k_i^2-k_\rho^2$, $z$ is the direction normal to the structure, and the subscript $i=1,2,3$ refers to the top (air), inner (PMMA), and bottom (Al$_2$O$_3$) dielectrics. In addition, $k_i$ denotes the wavenumber of the medium $i$ and $k_\rho = \beta - j\alpha$ is the complex wavenumber of the propagating plasmon. Note that we impose Im$[k_{z(1,2)}]<0$ in order to fulfill Sommerfeld's radiation condition and we assume that the stack width $W$ is much larger than the guided wavelength (i.e., $W>>1/k_\rho$). The supported even TM and odd quasi-TEM modes described by this dispersion relation can be accurately modeled using per-unit-length equivalent circuits\cite{Rana08, Diego13_mtt, Yoon14} (see Supplementary Note $2$).
\section{Author Contributions}
J.~S.~G. and J.~P.~C. conceived the idea of reconfigurable graphene stacks as well as their application in THz plasmonics. J.~S.~G. developed the theory and analyzed the measured data. C.~M. and A.~I. fabricated the devices. L.~S.~B. and A.~M. growth and transferred the graphene layers. S.~C. and J.~R. performed the THz measurements. J.~S.~G. wrote the manuscript (with comments from J.~P.~C and S.~C.). J.~P.~C. led the project.
\section{Competing financial interest}
The authors declare no competing financial interest.

\acknowledgement
This work is dedicated to the memory of Professor Julien Perruisseau-Carrier, of the École Polytechnique Fédérale de Lausanne (Switzerland), who
passed away unexpectedly while the manuscript was being prepared for publication.

This work was partially supported by the Swiss National Science Foundation (SNSF) under grant $133583$, by the European Commission
FP$7$ projects ``Grafol'' (Grant. No. $133583$), Marie-Curie IEF ``Marconi'' (ref. $300966$) and Marie-Curie ITN ``NAMASEN'', and by Ministerio de Economía y Competitividad, Spain, under grant CONSOLIDER CSD2008-00068.


\newpage
\section{Figures}

\end{document}